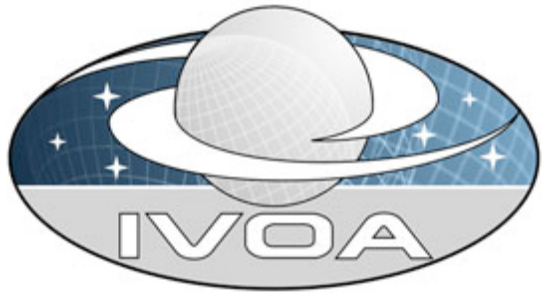

**I**nternational
**V**irtual
**O**bservatory
**A**lliance

# An IVOA Standard for Unified Content Descriptors
# Version 1.1

## IVOA Recommendation 2005-08-12

**This version:**
   http://www.ivoa.net/Documents/REC/UCD/UCD-20050812.html
**Latest version:**
   http://www.ivoa.net/Documents/latest/UCD.html
**Previous version(s):**
   http://www.ivoa.net/Documents/PR/UCD/UCD-20041026.html
**Editor(s):**
   S. Derriere, A. Preite Martinez, R. Williams
**Author(s):**

Sébastien Derriere (derriere@astro.u-strasbg.fr)
Norman Gray (norman@astro.gla.ac.uk)
Robert Mann (rgm@roe.ac.uk)
Andrea Preite Martinez (andrea.preitemartinez@rm.iasf.cnr.it)
Jonathan McDowell (jcm@cfa.harvard.edu)
Thomas Mc Glynn (Thomas.A.McGlynn@nasa.gov)
François Ochsenbein (francois@astro.u-strasbg.fr)
Pedro Osuna (Pedro.Osuna@esa.int)
Guy Rixon (gtr@ast.cam.ac.uk)
Roy Williams (roy@cacr.caltech.edu)

## Abstract

This document describes the current understanding of the IVOA controlled vocabulary for describing astronomical data quantities, called Unified Content Descriptors (UCDs).

The present document defines a new standard (named UCD1+) improving the first generation of UCDs (hereafter UCD1). The basic idea is to adopt a new syntax and vocabulary requiring little effort for people to adapt softwares already using UCD1.

This document also addresses the questions of maintenance and evolution of the UCD1+. Examples of use cases within the VO, and tools for using UCD1+ are also described.

## Status of This Document

This is a Recommendation. The first release of this document (An IVOA standard for Unified Content Descriptors version 1.1) was 2005-08-12.

This document has been produced by the IVOA UCD Working Group.

## Acknowledgments

This document is based on the W3C documentation standards, but has been adapted for the IVOA.

## Contents:

# 1  Scope of UCD

## 1.1  A Controlled Vocabulary for Astronomy

The Unified Content Descriptor (UCD) is a formal vocabulary for astronomical data that is controlled by the International Virtual Observatory Alliance (IVOA). The vocabulary is *restricted* in order to avoid proliferation of terms and synonyms, and *controlled* in order to avoid ambiguities as far as possible. It is intended to be flexible, so that it is understandable to both humans and computers. UCDs describe astronomical quantities, and they are built by combining words from the controlled vocabulary.

A UCD does not define the units nor the name of a quantity, but rather "*what sort of quantity is this?*"; for example **phys.temperature** represents a temperature, without implying a particular unit.

It would be possible to describe astronomical data quantities in a natural language such as English or Hungarian or Uzbek; however, it would be very difficult to expect a machine to 'understand' it in any sense. At the opposite extreme, there is an attempt within the IVOA to describe astronomical data in terms of a hierarchical data model, so that there is a place for everything, and everything is in its place. The UCD vocabulary falls between these extremes, and is intended to be understandable to both humans and computers.

## 1.2  Interoperability as a Goal

The UCD working group has tried to resist the temptation to allow the UCD syntax to be overly expressive. Every measurement in science has the possibility of essentially infinite description - the people, the instruments, the error analysis, the reasons, the funders, and so on. We have tried to find a way of organizing specifiers (words) so that it is easy to write simple software for machine use, but also possible to write better, more sophisticated software. We hope to build more sophisticated "intelligent" systems in the future.

The major goal of UCD is to ensure interoperability between heterogeneous datasets. The use of a controlled vocabulary will hopefully allow an homogeneous, non-ambiguous description of concepts that will be shared between people and computers in the IVO.

We hope in the future to put more semantic expressiveness into the UCD framework, but always keeping a pragmatic eye on those who would create and use the software that is to "understand" UCD.

## 2  UCD Syntax

A UCD is a string that contains textual tokens that we shall call **words**, which are separated by semicolons (;). A word may be composed of several **atoms**, which are separated by period (.) characters. The order of these atoms induces a hierarchy. Standard UCDs, which are validated by the IVOA, can start with the **ivoa:** namespace, but this namespace is optional. The use of namespaces, indicated by the presence of a colon in the word, is possible, but should be avoided as far as possible. They should be used only temporarily, for words that are not yet included into the vocabulary validated by the IVOA, and they should be replaced by the standard word as soon as it is validated. Section [4](#) describes a procedure for incorporation of new UCDs into the IVOA-approved list.

The character set that may be used in a UCD is the upper and lower-case alphabet, digits, hyphen, and underscore. The colon, semicolon, and period, are special characters as discussed above.

- There should be no space characters within a UCD. Space characters or tabs must be removed for a UCD to be well written.
- The UCD syntax is case-insensitive -- all uppercase characters should be converted to lowercase before parsing.
- It is ensured that for any two words at the same level in the hierarchy, one can't be a starting substring of the other. This means that xxx.abc and xxx.abd can exist, but not xxx.abc and xxx.abcd (the latter one having the former as starting substring).

The building blocks for UCDs are the *words* (like **phys.temperature**), not the *atoms* (like **temperature**). People trying to assign a UCD to an astronomical quantity should first describe in natural language *what the quantity is*, and then search the list of valid *words* for the best matching *words* in the UCD vocabulary. In most cases, one single word will be sufficient, and the UCD will simply be this word. Guidelines on how to combine several words are given in section [3.3](#). Some tools can help in automating this assignment (see appendix [C](#)).

Creation of new words is the responsibility of the UCD Scientific Board (see section [4](#)), and should occur when the vocabulary is missing some useful knowledge description. Atoms are only considered at this level (creation of a new word): once created, words become the basic elements from which UCD are built.

### 2.1  Examples of Legal Syntax

The following examples have legal UCD syntax:

- **pos.eq.ra;meta.main**
- **meta.id;src**
- **phot.flux;em.radio;arith.ratio**
- **PHot.Flux;EM.Radio;ivoa:arith.Ratio**

Notice that the last two UCDs are identical because of the case insensitivity and because the default namespace is optional.

## 2.2 Backus-Naur Form

```
<alpha>     ::=  a|b|c|d|e|f|g|h|i|j|k|l|m|n|o|p|q|r|s|t|u|v|w|x|y|z
                |A|B|C|D|E|F|G|H|I|J|K|L|M|N|O|P|Q|R|S|T|U|V|W|X|Y|Z
<digit>     ::=  0|1|2|3|4|5|6|7|8|9
<char>      ::= <alpha>|<digit>|-|_
<semicolon> ::== ;
<period>    ::== .
<colon>     ::== :
<word-component> ::= <alpha>|<digit>|<word-component><char>
<namespace-ref> ::= <word-component>
<word>      ::= <word-component>|<word><period><word-component>
<nword>     ::= <namespace-ref><colon><word>|<word>
<UCD>       ::= <nword>|<UCD><semicolon><nword>
```

Note: *A UCD is always case-insensitive.*

# 3  UCD Construction

## 3.1  From UCD1 to UCD1+

What makes UCD1 easy to use is that they are simple strings: they can be considered as a single word. But the immediate drawback, as it has been discussed many times, is that this implies creating many new UCD1 for only slightly different things.

> Consider the following list of 4 distinct UCD1 words:
>
> - **POS_EQ_RA_MAIN**
> - **POS_EQ_RA**
> - **POS_EQ_DEC_MAIN**
> - **POS_EQ_DEC**
>
> They reduce in fact to only 3 elements (**POS_EQ_RA**, **POS_EQ_DEC**, and **MAIN**), that could be combined to build the 4 fully-qualified terms.

***The idea of building UCDs by combining simple words makes the vocabulary less complex and more flexible*** (cf. "atomic UCDs" proposal by G. Rixon). The two questions that immediately arise are:

1. How do we define the simple words ?

2. How do we combine them to build fully-qualified UCDs ?

## 3.2 Defining Simple Words

There is no definitive answer to the first question, because selecting some terms for inclusion in the vocabulary and rejecting other terms is necessarily subjective. The only possible validation of the selected vocabulary is to check its ability to describe properly a wide range of real data.

There are two caveats for the definition of the list of words:

- The words must not be too simple/ambiguous (or we have the same inconveniences as natural language);
- The words must not be too specific (or we come back to the UCD1 drawbacks).

> In order to avoid ambiguities, **each word of the vocabulary will have an associated definition in plain text** (and possibly related keywords).
>
> Words like **source** or **type** should only be used in the vocabulary with a very clear definition, and restrict only to one meaning in the case of homonyms (**source** can a priori mean an object in the sky, a program code, a bibliographic reference, ...).
>
> For this reason, and also in order to group similar words, **words are composed of atoms**. The first atoms in a word generally help specifying the context, and help understanding the word without reading its definition (e.g. **pos.galactic.lat** is the latitude in galactic coordinates while **pos.ecliptic.lat** is the latitude in ecliptic coordinates).

## 3.3 Combining Simple Words

While it is possible to build a UCD as a combination of several simple words, **the primary word carries most of the meaning as to "what the quantity is"**. After the primary word, subsequent words are arranged by decreasing importance (see section 3.4 for some examples on ordering the words).

In the proposed scheme, UCDs are built by adding words from left to right, with each new word specifying/qualifying the combination to its left. The most important words when comparing two UCDs are the first ones (see appendix B) .

People or software that don't want to manage composed UCDs can use only the first word of the composed UCD (called the *primary* word). This word must give a first-order description of the quantity that is being described. It can be used as in UCD1, with the only change that the underscore (_) is to be replaced by a period (.) in the parsing (cf. section 2 for syntax of UCD1+).

> The choice of the primary word (when a complex element is to be described) should be guided by the answer to the question: "in *one word*, what is this element?".
>
> The units can give a hint to find the most appropriate primary word.
>
> ***One UCD describes one element, and if several elements (e.g. columns of a table) are present, the possible relationships between the elements are not used for attributing UCDs.***

**Example:** Consider a table containing 3 columns:

- A magnitude;
- A flag on this magnitude.
- An error on this magnitude;

The primary word for the first column should be **phot.mag**: the contents of this column is a number, and the semantic meaning of this column is well described by the word **phot.mag** (whose definition is *photometric magnitude*).

The contents of the second column is a flag (often it is a symbol, like a, b, or * that indicates, e.g., bad weather, unreliable values, ...). Therefore, the primary word should be **meta.code** (which means *code or flag*), because what is really described here is indeed a flag. The complete UCD could be written **meta.code;phot.mag**, to indicate that this flag applies to a magnitude. A simple parser could keep only the primary word of this UCD, and still have a reliable description of what it is. It could also ignore the order of all secondary words.

The content of the third column is an uncertainty, a measurement error. It can be expressed in magnitudes, but it is not a magnitude, so it is not correct to use **phot.mag** as primary word. One should use instead **stat.error** as the primary word, because the definition of this word corresponds precisely to the content of the column. The complete UCD could be written **stat.error;phot.mag**, to indicate that this error applies to a magnitude.

One could argue that these three columns are in fact related. This is correct, but it does not imply that the exact relation can be inferred from the UCD themselves. There are other expressive means to describe relationships between elements (e.g. use of <GROUP> tags in VOTable).

UCD1+ has been intentionally kept simple, using just simple combinations of words that describe elements. The idea is that ultimately, a UCD3 system, using RDF and/or ontologies, will allow a precise description of the relationships between elements. But this will lead to much more complex UCDs, that will most likely be no longer human-readable (or writeable). ***We hope however that most of the simple words that are defined in the UCD1+ vocabulary will be reusable in future evolution of UCDs.***

Appendix B describes how UCD1+ can be used in practice *now*, despite (or taking profit from) their simplicity.

**Examples of UCD1+ and how they are built:**

- *The maximum temperature of an instrument.* This is a *temperature*, so the primary word will be: **phys.temperature**. This temperature is that of an *instrument*, so we specify it next: **phys.temperature;instr**. And finally, we add a third word to indicate that this is the *maximum* value of a **phys.temperature;instr**, giving the final UCD:

  **phys.temperature;instr;stat.max**

- *The error on a magnitude measured in the V band.* The quantity is an *uncertainty*, so the primary word will be **stat.error**. This uncertainty applies to a *magnitude*, so we write **stat.error;phot.mag**. Then, we can specify the photometric band with another word, giving **stat.error;phot.mag;em.opt.V**.

In most cases, as shown in appendix A, one or two words are sufficient to form a UCD.

We can note that some of the words present in the vocabulary can not be used as primary words to describe a simple quantity (e.g., most of the words starting with **em.** that only describe a part of the electromagnetic spectrum). Such words that can not be used as primary will be flagged in the list of standard words, so that people or tools trying to assign UCD1+ can avoid errors.

## 3.4 Special Words Combinations

Some words of the vocabulary need to be used in combination with other words. These suggested combinations are described in the document defining the list of UCD1+ words, with the definition on the corresponding syntax flags. We list here a few construction rules that should resolve the most frequent problems in combining the words:

1. Words like **phot.mag**, **phot.flux**, **phot.count** are quantities that should be followed by one secondary word from the **em** branch, indicating in which part of the spectrum the quantity was measured. E.g. **phot.mag;em.IR.J** represents a magnitude measured in the J band.

   This two-words combination should then be treated as a single block in the case where other words are to be added to this UCD.

2. **phot.color** should be followed by two words from the **em** branch, to express what bands were used in the computation of the color. For example, **phot.color;em.opt.B;em.opt.V** in that order represents the *B-V* color (while **phot.color;em.opt.V;em.opt.B** would mean *V-B*).

3. Some words (**meta.main**, **meta.modelled**, **stat.mean**, **stat.median**, **stat.min**, **stat.max**) are qualifiers, and it is recommended to place them after the quantity they qualify. For example, **phys.temperature;stat.max** or

**phot.mag;em.IR.J;meta.modelled** are valid UCD (the latter illustrating points 1 and 3).

4. **arith.ratio** is used to represent a ratio between two quantities represented by the same UCD. For example a temperature ratio $T_1/T_2$ is represented by **phys.temperature;arith.ratio**. It can not be used for ratios between different quantities (e.g. mass-to-light ratio).

   Similarly, **arith.diff** is used to represent a subtraction between two quantities represented by the same UCD.

## 4 UCD Boards

### 4.1 Creation of a Scientific Board for New UCD Words

The inclusion of new UCD words must be a flexible but controlled process. A Scientific Board will study new UCD requirements and, within a given period of time, give an answer as to whether a new UCD must or must not be included in the UCD standards.

The use of "mission-specific" namespaces has been addressed in many occasions, and we believe that namespaces should be avoided as much as possible. There has been an exercise in revising the VOX words for the SIAP protocol and trying to assign existing UCDs to them, or proposing new UCD words for the non-existing ones.

The responsibility of the Board consists in studying the cases where a UCD word is proposed and to figure out whether the proposed word should be accepted or rejected, and in case of rejection, recommending the closest existing word that should be used.

In case a new word is accepted into the main tree, an internal procedure is established so that the new UCD becomes live after a proper internal new release in a short period of time. It should be agreed whether this Board would study the proposed cases in an "on demand" basis or would collect requests and study them on a periodic basis.

This Scientific Board is composed of astronomers with broad experience in different subdisciplines as well as data providers. They should have the experience and the resources to maintain the UCD system.

The Board's mission is only to maintain and improve the list of words (see 4.5). If major modifications were to be made to the UCD structure, the way they are built, or their goal, then the reference document would have to go through the whole IVOA validation process.

### 4.2 A Procedure to Request New UCD Words

A procedural document should be created to make it easy to a user to ask for a new UCD word and to understand the implications of doing so. This document would address:

- the contact point to ask for new UCD
- the life-cycle of the process of asking for a new UCD
- when and how a new UCD becomes live
- what to do if a UCD is rejected

These actions should be supported by tools like an automatic form that is filled in and sent to the Scientific Board, giving an answer back to the user acknowledging the request, and giving a time estimate for an answer.

A first interface for requesting new UCD words has been set up at: http://cdsweb.u-strasbg.fr/UCD/cgi-bin/comment/ucdComments.

## 4.3 Creation of a Technical Board

There should be tools available for the user to check for the existence of UCDs, etc. Some of these tools exist already (see appendix C), and they are good candidates to become the sort of "official" tools for the UCD standards. A Technical Board should be established to decide which tools are really necessary to make the UCD work feasible and as easy as possible for the user. This Board will be mainly in charge of writing proper requirements for the tools.

## 4.4 Contact Point for UCD Issues

It is useful to have a contact point to which all UCD related matters can be addressed. This contact point could be a web address devoted explicitly to that in the context of the VO, a properly organized web place, where all the tools would be available, as well as all documents and procedures for creation of new UCD words, etc., with practical examples and the like.

Currently, most of UCD-related material is available at this address :
http://cdsweb.u-strasbg.fr/UCD/

## 4.5 List of Valid Words and UCDs

The list of valid words is not included in this document: it is described in a separate IVOA document: http://www.ivoa.net/Documents/latest/UCDlist.html

The Scientific Board keeps the list under configuration control. Tools are provided for people using UCDs, to allow easy validation, and to ensure compliance of the words with the latest version of the list (see the *upgrade* method in appendix C).

# A  Transition from UCD1 to UCD1+

Services or protocols that already use UCD1 can evolve to use UCD1+ with little extra work. This is because, in most cases, they use standard elements that can be easily expressed with simple combinations of words.

The flexibility of UCD1+ can also be exploited. For example, the Cone Search currently expects the use of the UCD1 **POS_EQ_DEC_MAIN**. This element would now be written **pos.eq.dec;meta.main**. The **meta.main** word is in fact only useful when there are several values of declination in the same dataset. If there is only one value of a declination, it can be described by **pos.eq.dec**, and a flexible matching function could indicate that this UCD is compatible with the required **pos.eq.dec;meta.main** (cf appendix B).

The definition of a new list of words is also the occasion to describe in a homogeneous way elements that do not exclusively come from the UCD1:

- a description of FITS keywords coming from different image headers adds useful words to the vocabulary;
- the VOX words that are used for the SIAP protocol can be replaced by combination of UCD1+ words. Only a few words not used in the vocabulary defined by UCD1 are required;
- the atomic and molecular data description is also underway, to provide a homogeneous description of large reference databases in this domain (BASECOL, ...).

# B  Use Cases

## B.1  Database Access and UCDs: Translation Layer

UCDs will be used in practice for *exchanging* information using a controlled vocabulary. They are used in the VOTable standard to attach a standard description to table column names, for example. The data providers do not need to change the internal descriptions of their existing databases. Nor is it required that people building from scratch a new VO-compliant service use UCDs in the core of their system.

What is needed for interoperation with other systems is a "*translation layer*" that is able to associate UCDs to the parameters that are used internally, so that the output of the service contains a standard description that can be interpreted by other VO services.

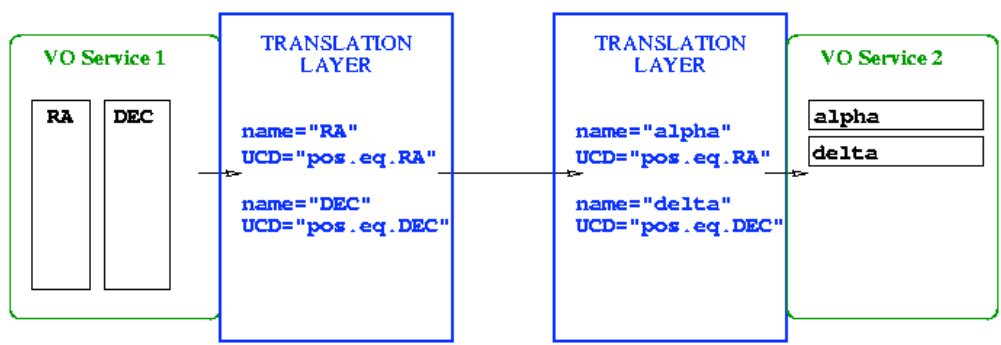

**Figure 1:** Services use UCD to exchange information. A translation

11 of 14

layer is used to interpret the internal description in terms of UCD.

In Fig. 1, a first VO service describes internally the right ascension and declination with names RA and DEC. For sending data to another service expecting right ascension and declination as an input, it uses a translation layer to attach UCDs to its parameters. The second service also has a translation layer that can interpret UCDs into its own parameters.

The mapping done by the translation layer can be done using XML files. For the second service above, the quantities corresponding to UCDs **pos.eq.ra** and **pos.eq.dec** are to be found in the database table Obs-Table, which has column names alpha and delta:

```xml
<?xml version='1.0'?>
<!DOCTYPE ucdToDb SYSTEM 'ucdToDb.dtd'>
<ucdToDb>
    <ucd name="pos.eq.RA" table="Obs-Table" col="alpha" />
    <ucd name="pos.eq.DEC" table="Obs-Table" col="delta" />
    <ucd name= ... />
</ucdToDb>
```

## B.2  UCDs in VO Tools

There are already applications that use UCD1 to manipulate or display some data to the user, or to find required fields (VOPlot, Filters in Aladin, etc.). These applications can continue to function, using the primary UCD1+ word only.

With UCD1+, it is possible to be more flexible, and to find the "most appropriate" element in a dataset.

Consider a tool that expects to find a field with UCD **pos.eq.ra;meta.main**. Using a custom matching function (see appendix B) to analyze the contents of a VOTable file, this tool could consider that **pos.eq.ra** matches in the absence of **pos.eq.ra;meta.main**, and pick that column as the expected one.

## B.3  UCDs in a Registry

A registry can contain descriptions of catalogues, with the associated UCDs. The benefit of having access to the contents in terms of UCDs is that it is possible to explore the contents of a catalogue more extensively than with simple keywords.

For example, a catalogue dedicated to very accurate measurement of proper motions and parallaxes will certainly put keywords for these, but it might also contain a column that measures a radial velocity. With UCDs assigned, this column could be identified and the catalogue selected for someone searching for radial velocities, even if this is not the primary goal of the catalogue.

It is however not necessary to describe *every* element of a dataset by UCDs. Only the most relevant columns need to have UCDs attached to them. Parameters used for internal processing by a service do not need to have UCDs attached.

Consider the catalogue above described with UCDs in a registry. A query by UCD allows one to locate this catalogue and find that it contains radial velocities.

Once the resource is located, one can then send a query to this resource, either on its specific parameters or again using UCDs.

Because UCD1+ have a more flexible syntax, it is possible to make some kind of fuzzy search, with the help of matching functions (see appendix B) in the case of the search in a registry.

The different possible levels of granularity in the description allow more interoperability.

## C  Software and Services

Tools and services designed for UCD1+ are available online (and also as Web Services) at http://cdsweb.u-strasbg.fr/UCD/.

Those currently available are:

- **UCD Browser:** This tool allows a dynamic view of the tree of UCD1+, either as a single text file, or as a Javascript-enabled tree-browser.
- **translate:** this tool translates a UCD1 into the default corresponding UCD1+.
- **explain:** this tool returns a textual description of the meaning of a given UCD1+.
- **validate:** this tool checks if a UCD1+ is correctly written.
- **upgrade:** this tool transforms a deprecated UCD1+ into the recommended expression according to the latest version of the list of words.
- **assign:** This service returns a UCD1+ corresponding to a plain text description.

The UCD Technical Board (see section 4.3) is responsible for proposing and designing new tools.

## D  Changes from previous versions

### D.1  Changes from v1.06

- The sections on the Matching Function and Consequences for VizieR have been moved into separate notes.
- Following RFC discussion, an additional constraint on UCD syntax has been added (one UCD should not be substring of another at the same level in the hierarchy).
- The discussion on how words are combined (3.3) has been revised
- The section on the UCD steering comitee has been updated to acknowledge decisions made at the Kyoto interoperability meeting: creation of a Scientific Board and a Technical Board.
- Section 7 has been updated to describe new reference UCD1+ tools
- Abstract slightly changed.
- Section 5 has been renamed and moved to Appendix A.

- Sections 6 and 7 have been moved to Appendix B and C, respectively.
- The document has been edited to remove informal language.

## D.2 Changes from v1.05

- A sentence was added at the end of the abstract to describe section 9.
- Status of this document section changed to reflect PR status, and location of RFC page on IVOA wiki.
- Last sentence of section 4.2 removed.
- main changed to **meta.main**, section 5, paragraph 2.
- Last paragraph of section 9.1 p.13 changed.
- URL of request for new words moved from section 9.4 to 9.2.

## D.3 Changes from v1.03

- Document moved from IVOA Working Draft to Proposed Recommendation.
- Addition of 2 paragraphs in section 2 (p. 4), to highlight that basic elements in UCD construction are words, not atoms, and make clear the atom/word/UCD distinction.
- Section 3.4 was added to discuss special combinations of words. This helps describing the order in which words are arranged in UCD that are not a single word. A link from section 3.3 to section 3.4 has also been added.
- The exact numbers of words and UCD combinations have been replaced by approximate values in section 4.2, as these are subject to small changes.
- Section 8 has been reorganized.
- A paragraph was added at the end of section 9.1 to precise the role of the UCD board.
- URLs have been added to section 9.3.
- The URL in section 9.5 has been changed from http://vizier.u-strasbg.fr/UCD/lists/ to http://cdsweb.u-strasbg.fr/UCD/
- Number of typos corrected.